\documentclass{ws-procs9x6}

\newcommand{\imag}{\mbox{i}\ }
\newcommand{\cal }{\mathcal }

\begin{document}

\title{What we are learning about the Quark Structure of Hadrons from Lattice QCD \footnote{\uppercase{T}his work is supported by \uppercase{DOE} cooperative ageement  \uppercase{DE-FC}02-94 \uppercase{ER}40818.}}

\author{J.~W. Negele}

\address{Center for Theoretical Physics, \\
Massachusetts Institute of Technology, \\ 
77 Massachusetts Ave. \\
Cambridge, MA 02139, USA\\ 
E-mail: negele@mitlns.mit.edu}

\maketitle

\abstracts{Developments in lattice field theory and computer technology have led to dramatic advances in the use of lattice QCD to explore the quark structure of hadrons. This talk will describe selected examples, including structure functions, electromagnetic form factors, the nucleon axial charge, the origin of the nucleon spin, the transverse structure of the nucleon, and the nucleon to Delta transition form factor.}

\section{Introduction}



Lattice field theory is coming of age as an essential tool for exploring hadron structure. In addition to the prospect of precisely
calculating the experimentally observable properties of the nucleon
from first principles, it also offers the deeper opportunity of
obtaining insight into how QCD actually works in producing the rich
and complex structure of hadrons.  Beyond simply calculating numbers,
we would like to answer basic questions of hadron structure. For example, what are the dominant components of the nucleon wave function? How does the total spin of the nucleon arise from the spin and orbital angular momentum of its quark and gluon constituents?  How does the nucleon quark and gluon structure produce the observed scaling behavior of form factors?  What is the transverse,
as well as longitudinal structure of the nucleon light-cone wave
function?  As the quark mass is continuously decreased from a world in
which the pion mass is 1 GeV to the physical world of light pions, how
does the physics of the quark model and adiabatic flux tube potentials
evolve into the physics of chiral symmetry breaking, where instantons,
quark zero modes, and the associated pion cloud play a dominant role? What is the role of diquarks in conventional hadrons and exotic states such as pentaquarks? As discussed below, contemporary lattice calculation are beginning to provide insight into these and other fundamental questions in hadron
structure.

\begin{figure}[t]
\begin{center}
 \vspace*{-0.2cm}
 \includegraphics[scale=0.60,clip=true,angle=0]{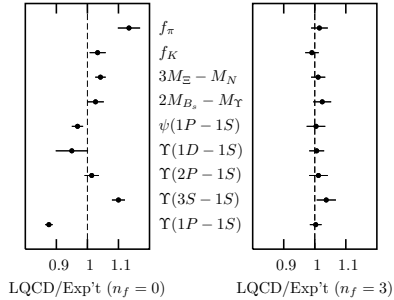}
  \vspace*{-0.1cm}
 \caption{First principles calculations using improved staggered quarks of selected decay constants and hadron mass differences   
showing agreement with experiment at the level of a few percent in full QCD (right) and showing discrepancies up to 10 percent in the quenched approximation, which ignores dynamical quark-antiquark excitations, (left). }
 \label{fig:GoldPlated}
 \end{center}
   \vspace*{-.5cm}
\end{figure}

 It has taken the first thirty years since Wilson's seminal formulation of QCD on a lattice\cite{Wilson:1974sk} to develop the theoretical techniques and computer technology for  quantitative numerical solution of QCD. Reference \raisebox{-.12cm}{\cite{Negele:1997bi}}  provides an elementary 
introduction for non-specialists.  The basic practical problem is that one must address the triply 
 demanding limits of a small lattice spacing, a large physical volume, and a small quark mass.   Instead of referring to the unobservable bare quark mass, since $ m_q \propto m_\pi^2 $, it is convenient to express the quark mass dependence of observables by their dependence on  $ m_\pi^2 $. To include the essential physics of the pion cloud, the box size must be large compared to the pion Compton wavelength, $ m_\pi^{-1} $, 
and the ultimate computational cost of a full QCD lattice calculation including dynamical sea fermions  turns out to have mass dependence $m_\pi^{-9}$. Hence, in the past, most full QCD calculations have been relegated to what I call the ``heavy pion world", where $m_\pi \ge $ 500 MeV, and it was conventional to perform theoretically unjustifiable linear extrapolations in  $ m_\pi^2$ to obtain first estimates of physics in our world with 140 MeV pions. Recently, using computationally economical staggered sea quark configurations with the so-called asqtad improved action generated by the MILC collaboration\cite{Orginos:1999cr,Orginos:1998ue},  a number of heavy quark observables have been calculated in the chiral regime of light pions and extrapolated using theoretically motivated chiral perturbation theory. Figure~\ref{fig:GoldPlated} shows impressive agreement with experiment at the level of a few percent of decay constants and mass splittings using this theory in full QCD, and also indicates how the quenched approximation, which omits quark excitations from the Dirac Sea, introduces discrepancies at the 10\% level\cite{Davies:2003ik}. Results discussed in this talk will include calculations in the heavy pion world with SESAM full QCD configurations\cite{Eicker:1998sy} using Wilson quarks or with quenched Wilson quarks, and some initial hybrid calculations in the chiral regime of light pions using MILC staggered sea quark configurations\cite{Orginos:1999cr,Orginos:1998ue} with domain wall valence quarks.

\section{Nucleon Structure}

I will briefly discuss the experimental observables that are calculable on the lattice and describe selected recent results. More details may be found in a recent review\cite{Negele:2004iu},  recent publications of our group\cite{Dolgov:2002zm,Hagler:2003jd,:2003is,Detmold:2001jb} and of the QCDSF collaboration\cite{Gockeler:2004vx,Gockeler:2003jf}.  

Since asymptotic freedom renders QCD corrections to high energy scattering small and calculable, high energy lepton scattering provides precise measurements of matrix elements of the light-cone operator\\ 
{\small
$ {\cal O}(x) \!=\!\int \!\frac{d \lambda}{4 \pi} e^{i \lambda x} \bar
  \psi (\frac{-\lambda n}{2})\!\!
  \not n {\cal P} e\!^{-ig \int_{-\lambda / 2}^{\lambda / 2} d \alpha \, n
    \cdot A(\alpha n)}\!
  \psi(\frac{\lambda n}{2})$}, 
where $n$ is a unit vector along the light-cone. Expanding  ${\cal O}(x) $ in local operators via the operator
product expansion generates the tower of twist-two
operators,\\
$ {\cal O}_q^{\lbrace\mu_1\mu_2\dots\mu_n\rbrace} = {\bar \psi}_q
   \gamma^{\lbrace\mu_1} \imag{D}^{\mu_2} \dots
  \imag{D}^{\mu_n\rbrace} \psi_q$,
whose matrix elements can be calculated in lattice QCD.


\begin{figure}[t]
\begin{center}
 \includegraphics[scale=0.45,clip=true,angle=0]{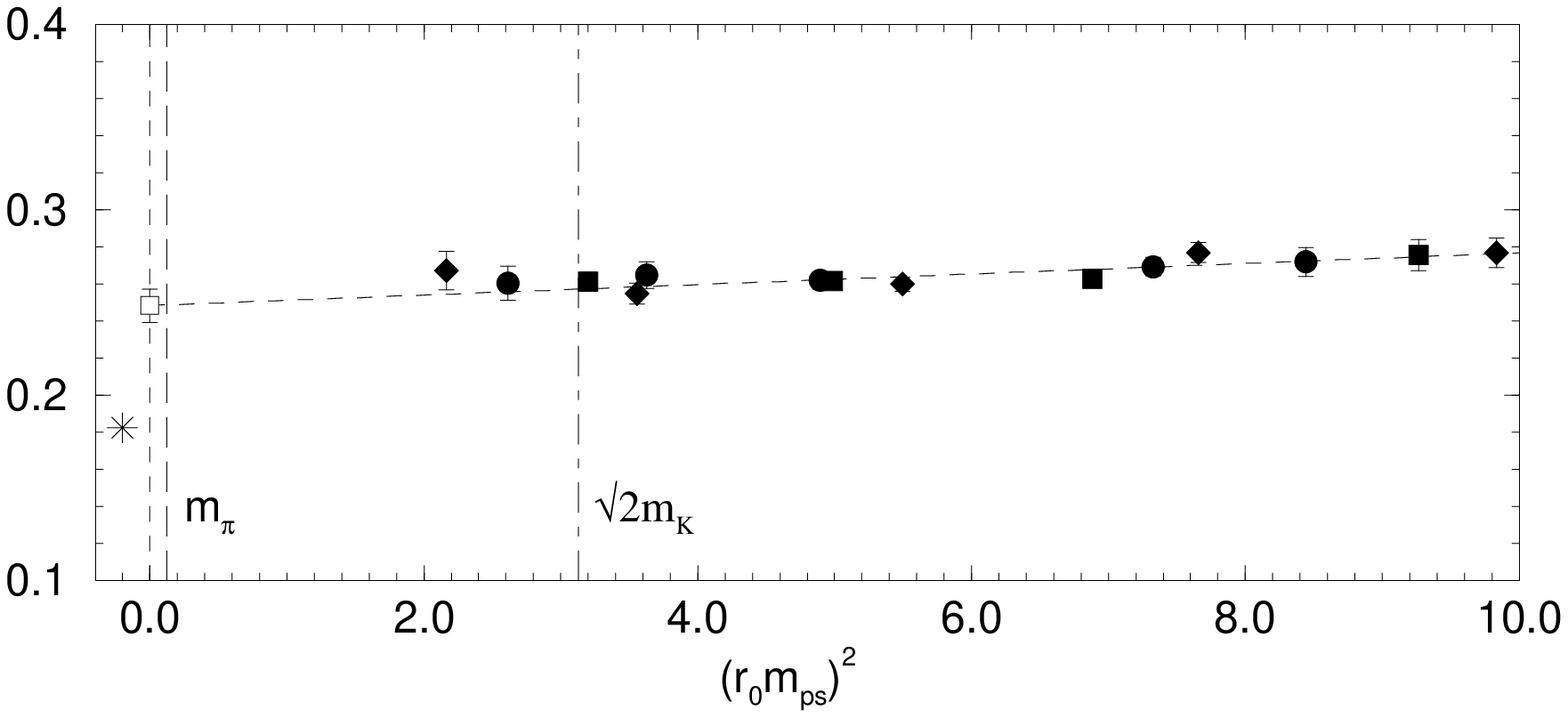} \\[.2cm]
 \includegraphics[scale=0.35,clip=true,angle=0]{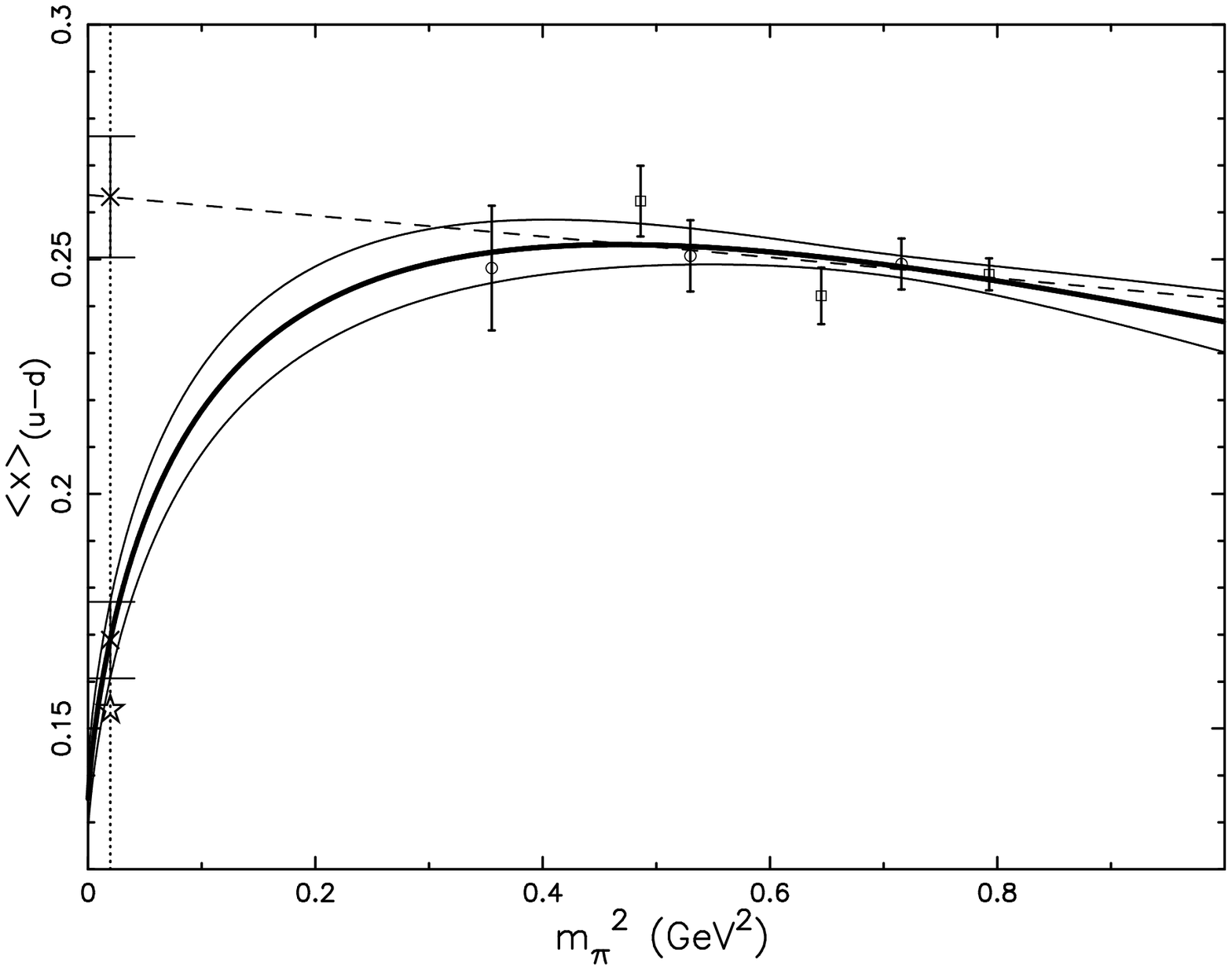}
 \caption{Linear extrapolation of unquenched calculations of the nonsinglet quark momentum fraction, $\langle x \rangle$, (top) and a chiral extrapolation of full QCD (bottom) as described in the text. }
 \label{fig:momfract}
 \end{center}
   \vspace*{-.1cm}
\end{figure}

The familiar quark distribution $q(x)$ specifying
the probability of finding a quark carrying a fraction $x$ of the
nucleon's momentum in the light cone frame is measured by the
diagonal nucleon matrix element, $ \langle P |{\cal O}(x) | P
\rangle = q(x) $,
and the diagonal matrix element $ \langle P | {\cal
  O}_q^{\lbrace\mu_1\mu_2\dots\mu_n\rbrace} | P \rangle$ specifies the
$(n-1)^{th}$ moment of the quark distribution, $\int
dx\, x^{n-1} q(x) $.
 Analogous expressions in which the 
twist-two operators contain an additional $\gamma_5$ measure moments of the
longitudinal spin density, $\Delta q(x)$.
 The generalized parton distributions  $ H(x, \xi, t)$ and  $ E(x, \xi, t)$  \cite{Muller:1994fv,Ji:1997ek,Radyushkin:1997ki,Diehl:2003ny} are
measured by off-diagonal matrix elements of the light-cone operator\\
$
\langle P' |{\cal O}(x) | P \rangle\! =  \!\langle\!\langle \not \! n \rangle\!\rangle
H(x, \xi, t) \nonumber  +  \frac{i \Delta_\nu} {2 m}  \langle\!\langle \sigma^{\alpha \nu} n_{\alpha} \rangle\!\rangle
E(x, \xi, t),
$
where
$\Delta^\mu = P'^\mu - P^\mu$, $ t  = \Delta^2$, $\xi = -n \cdot \Delta /2$, and
$\langle \!
\langle \Gamma \rangle \! \rangle = \bar U(P') \Gamma U(P)$ for Dirac spinor $U$.
Off-diagonal matrix
elements of the tower of twist-two operators
$ \langle P' | {\cal
  O}_q^{\lbrace\mu_1\mu_2\dots\mu_n\rbrace} | P \rangle$ yield moments of the
generalized parton distributions, which in the special case of $\xi$ =
0, are
$
 \int dx\, x^{n-1} H(x, 0, t)  =    A_{n, 0}(t)$, $ \int dx\, x^{n-1} E(x, 0, t) =  B_{n, 0}(t),$
where  $ A_{n, i}(t)$ and $B_{n, i}(t)$ are referred to
as generalized form factors (GFF's).  I will now address several special cases of these general expressions.

\begin{figure}[t]
\begin{center}
\vspace*{-0.8cm}
 \includegraphics[scale=0.30,clip=true,angle=0]{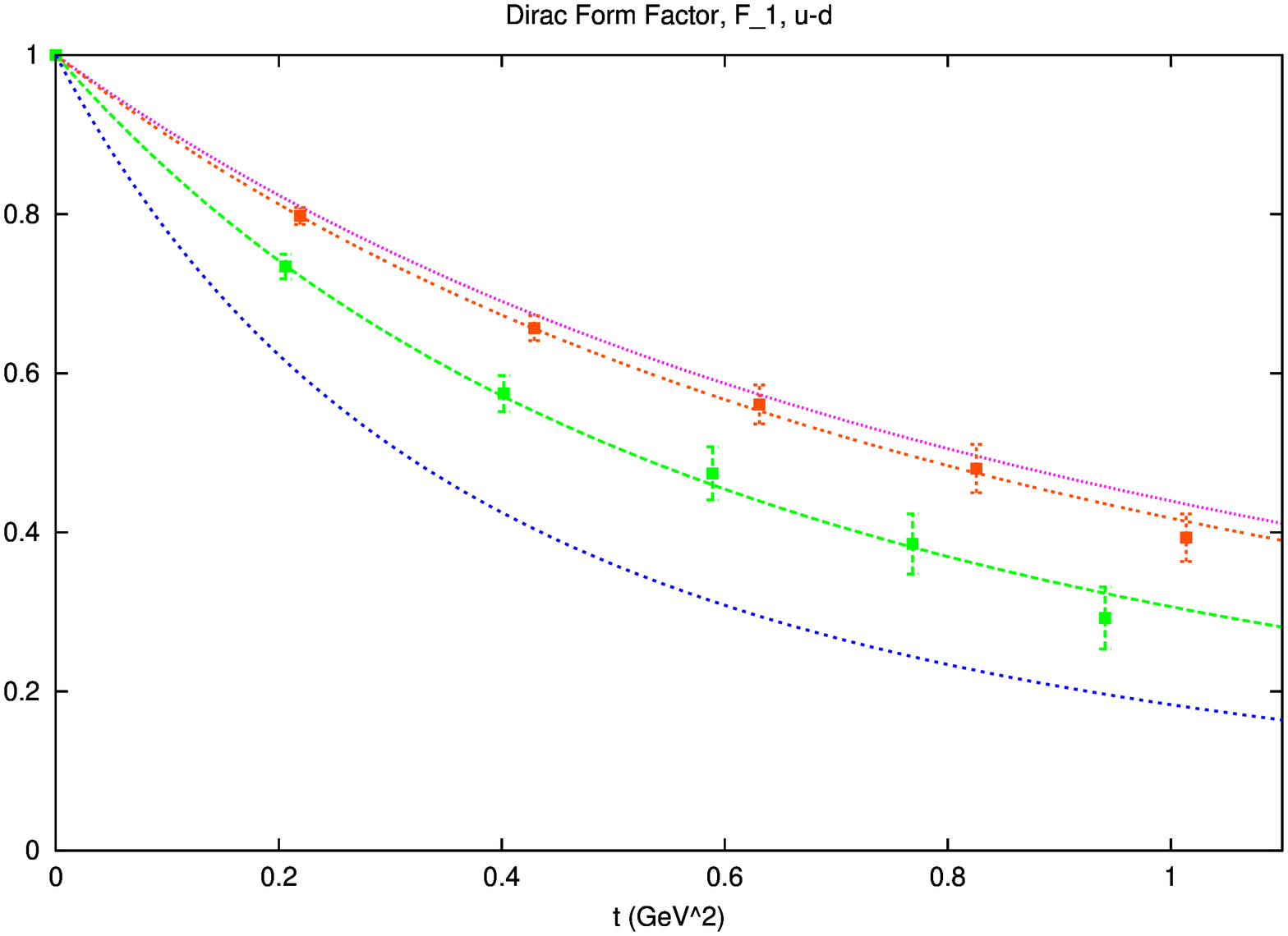}\\[-1.0cm]
 \includegraphics[scale=0.30,clip=true,angle=0]{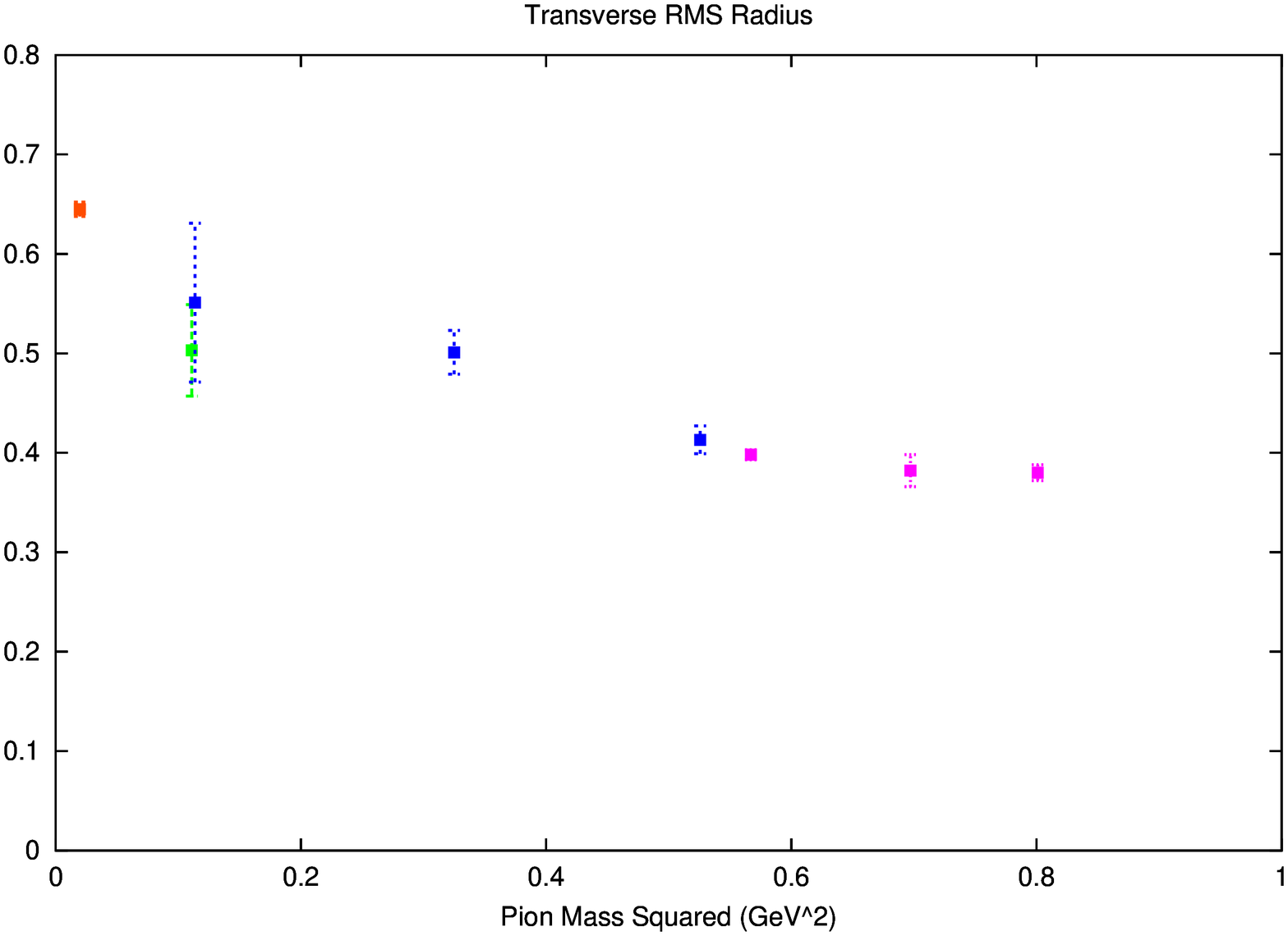}
   \vspace*{-0.6cm}
 \caption{Electromagnetic form factor $F_1$ of the proton (top panel) showing, in decreasing order, lattice results for pion masses 744, 725, and 570 MeV, and experiment.  The bottom panel shows the transverse rms radius corresponding to the slope of $F_1$ as a function of $m_\pi^2$ and the experimental result.}
 \label{fig:F1}
 \end{center}
  \vspace*{-.2cm}
\end{figure}

\smallskip
\noindent {\bf Quark momentum fraction}  \quad The quark momentum fraction, $\langle x \rangle_{q} = \int dx x q(x)  \propto \int dx \bar q \gamma^\mu D^\nu q $,  is particularly interesting, because it reflects the fact that a large fraction of the momentum is carried by gluons rather than quarks.  Figure~\ref{fig:momfract} shows the flavor nonsinglet difference between the momentum fraction of up and down quarks, which has no contributions from presently uncalculated disconnected diagrams and thus may be compared directly with experiment. The top plot shows that naive linear extrapolation of quenched calculations\cite{Gockeler:2002ek} yields a result 50\% higher than experiment. The full QCD calculations at the bottom have been extrapolated using the functional form $a [1-\frac{3g_a^2 + 1 ) m_\pi^2}{(4 \pi f_\pi)^2} ln(\frac{m_\pi^2}{m_\pi^2 + \mu^2})] + b m_\pi^2$, where $ \mu $ is a  phenomenological parameter joining the calculations in the heavy pion world to the light quark regime, which has the leading behavior described by chiral perturbation theory.  Although it is still a future challenge  to extend calculations of the momentum fraction into the chiral regime and observe this rapid turnover from first principles, we will see below two observables that have been successfully calculated in this regime and agree well with experiment.

\begin{figure}[t]
\begin{center}
 \vspace*{-0.6cm}
\hspace*{-.5cm} \hbox{ \includegraphics[scale=0.20,clip=true,angle=0]{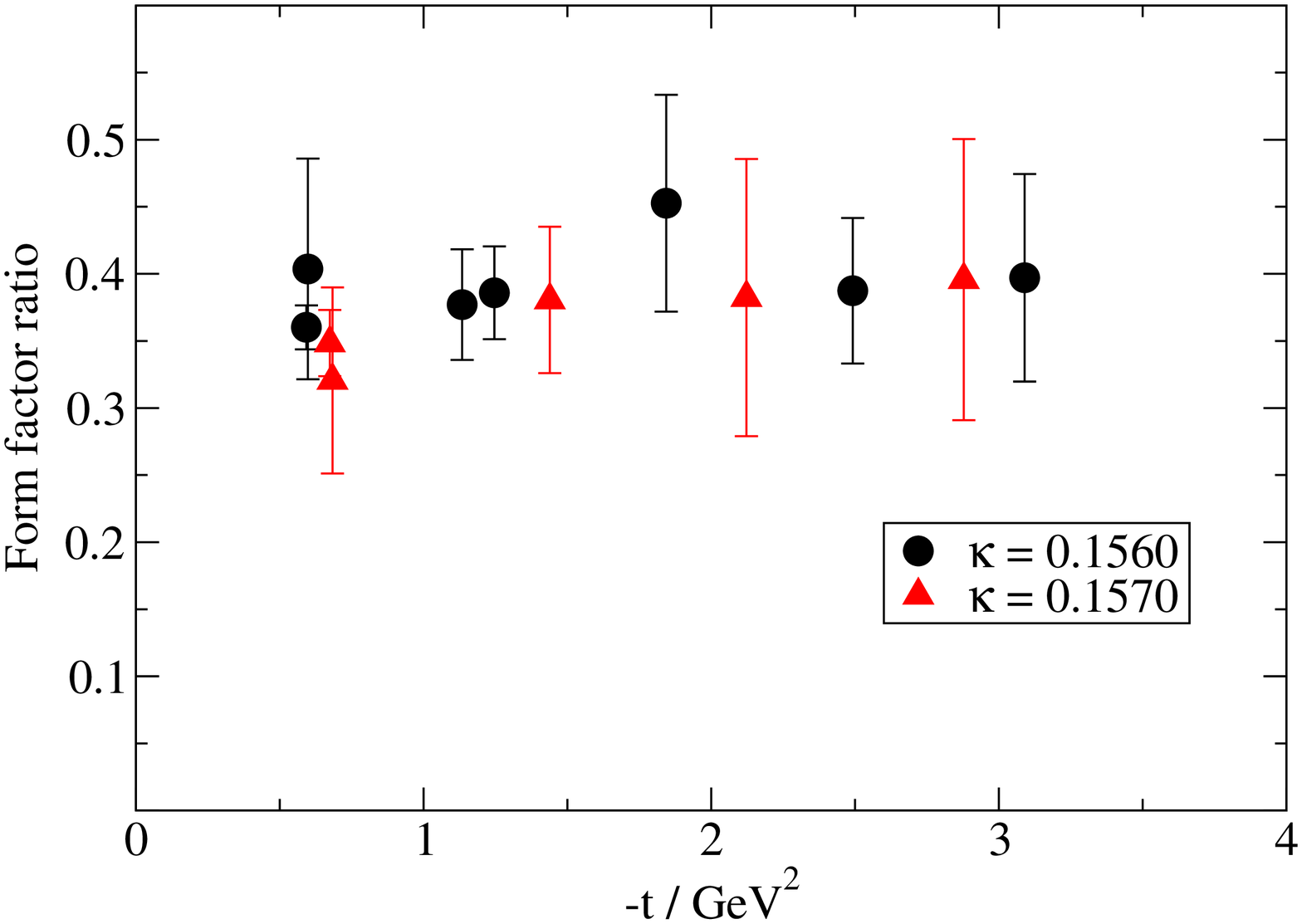} 
 \hspace*{-.5cm} \includegraphics[scale=0.43,clip=true,angle=0]{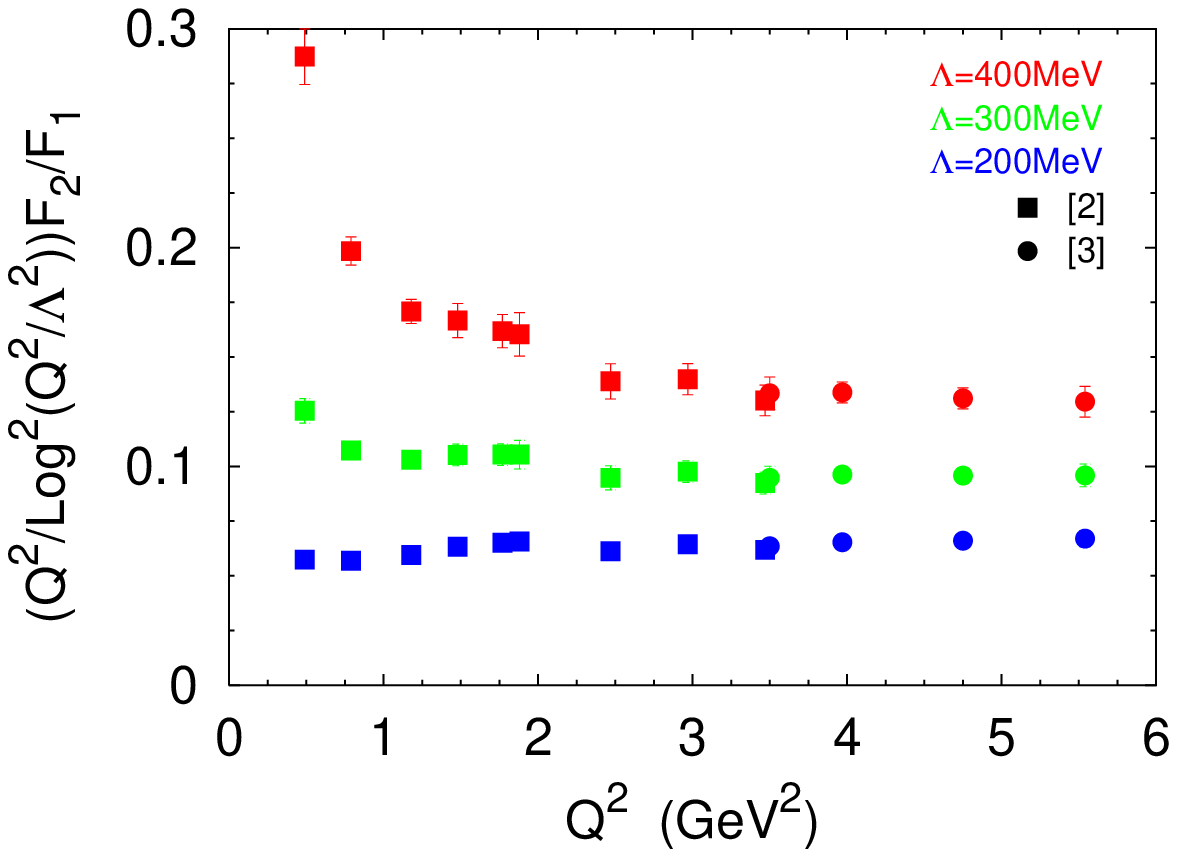}}
 \caption{Lattice results for $[Q^2 F_2]/[log^2(Q^2/\Lambda^2) F_1] $ for $m_\pi$ = 744 and 897 MeV as a function of momentum transfer (left panel) and from experiment (middle curve on right panel) for $\Lambda$ = 300 MeV.}
 \label{fig:F2/F1}
 \end{center}
   \vspace*{-.2cm}
\end{figure}

\smallskip
\noindent {\bf Electromagnetic form factors} \quad The electromagnetic form factors $F_1$ and $F_2$, corresponding to $A_{10}$ and $B_{10}$ defined above, characterize the spatial distribution of charge and current at low momentum transfer and the ability of a single quark to absorb a large momentum transfer and remain in the ground state. Figure~\ref{fig:F1} shows $F_1$, which specifies the Fourier transform of the transverse charge distribution in the infinite momentum frame, for Wilson fermions at $m_\pi$ = 744 MeV, and the hybrid combination of staggered sea and domain wall valence quarks at $m_\pi$ = 725 and  570 MeV. Note that the two heavy quark calculations are quite consistent, substantiating the equivalence of the two calculations and that the slope decreases toward the experimental result, corresponding to the increase in the spatial extent as the size of the pion cloud increases. The bottom plot shows that the transverse rms charge radius increases smoothly as the mass decreases, and is heading toward the experimental result.

One of the early successes of perturbative QCD was the understanding
of how the short range quark structure of a hadron governs the
behavior of exclusive processes at large momentum transfer. However,
whereas simple counting rules suggested that $F_2 \sim F_1 / Q^2$,
experimental data from JLab\cite{Gayou:2001qd} show that $F_2$ falls off much
more slowly. Theoretically, it has recently been shown
\cite{Belitsky:2002kj} that the next to leading order light cone wave
function yields $F_2 \sim F_1 {\log^2 (Q^2 / \Lambda^2)} / {Q^2}$,
and the agreement between this prediction with $\Lambda$ = 0.3 GeV and the JLab data is shown
in the right panel of Fig.~\ref{fig:F2/F1}.
 Since the short range quark structure dominates this physics, it is reasonable to expect that omission of the pion cloud in the heavy pion world should not destroy the qualitative behavior.  Indeed, our lattice results\cite{Negele:2004iu} plotted  in the left panel of Fig.~\ref{fig:F2/F1} for the value $\Lambda$ = 0.3 GeV yields excellent agreement with the $Q^2$ behavior of the experimental data.

\begin{figure}[t]
\begin{center}
 \vspace*{-0.6cm}
\hspace*{-1cm}  \includegraphics[scale=0.30,clip=true,angle=0]{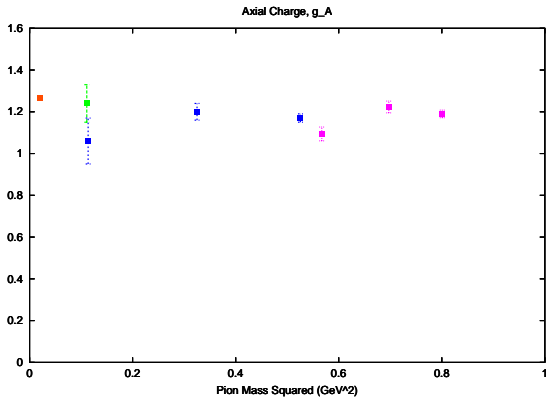}
  \vspace*{-0.8cm}
 \caption{Nucleon axial charge, $g_A$, calculated in spatial box sizes 1.5 fm (right three error bars), 2.6 fm (next two points and lowest error bar at left) and 3.5 fm (top error bar at left), compared with experiment (data point at far left). }
 \label{fig:gA}
 \end{center}
 \vspace*{-.2cm}
\end{figure}
\smallskip
\noindent {\bf Nucleon axial charge}  \quad The nucleon axial charge,  $g_A = \langle 1 \rangle_{\Delta q} = \int dx  \Delta q(x)  \propto \int dx \bar q \gamma^\mu \gamma^5 q $, that enters into $\beta $ decay, is particularly sensitive to finite volume effects that reduce the contributions of the pion cloud.  Figure~\ref{fig:gA} shows results in a series of three different box sizes. In each case, as the quark mass is decreased in a given box, the pion eventually stops fitting in the box and $g_A$ correspondingly artificially decreases. However, the locus of points measured in the largest boxes at each mass smoothly approaches the experimental result at the far left, providing a striking success of full QCD in the chiral regime.

\smallskip
\noindent {\bf Proton spin} \quad
In the nonrelativistic quark  model, the total proton spin of $1/2$ arises trivially from adding the spins of the three valence quarks, and the so-called spin crises arose when deep inelastic scattering measurements of the lowest moment of the spin-dependent structure function, $\Delta \Sigma = \langle 1 \rangle_{\Delta u} + \langle 1 \rangle_{\Delta d}$, indicated that only of the order of 30\% of the nucleon spin arises from quark spins. Hence, it is interesting to use the lattice to study how the angular momentum decomposition evolves as the pion mass is decreased from the heavy, quark model, limit to its physical value.

\begin{figure}[t]
\begin{center}
 \vspace*{-1.4 cm}
\hspace*{-1.0 cm} \hbox{ \includegraphics[scale=0.225,clip=true,angle=270]{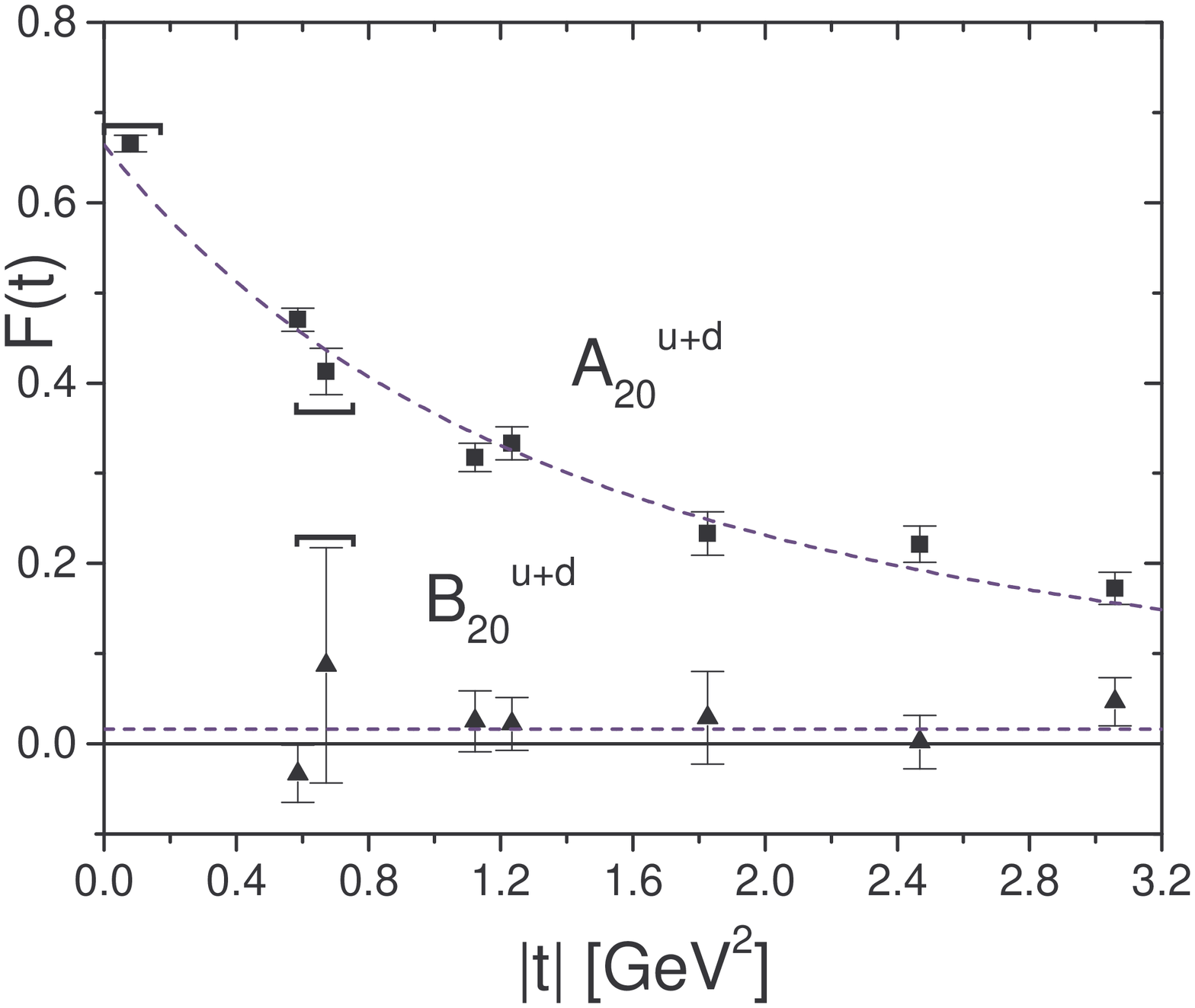}
 \hspace*{-1.0cm} \raisebox{-5.4cm}
 {\includegraphics[scale=0.33,clip=true,angle=0]{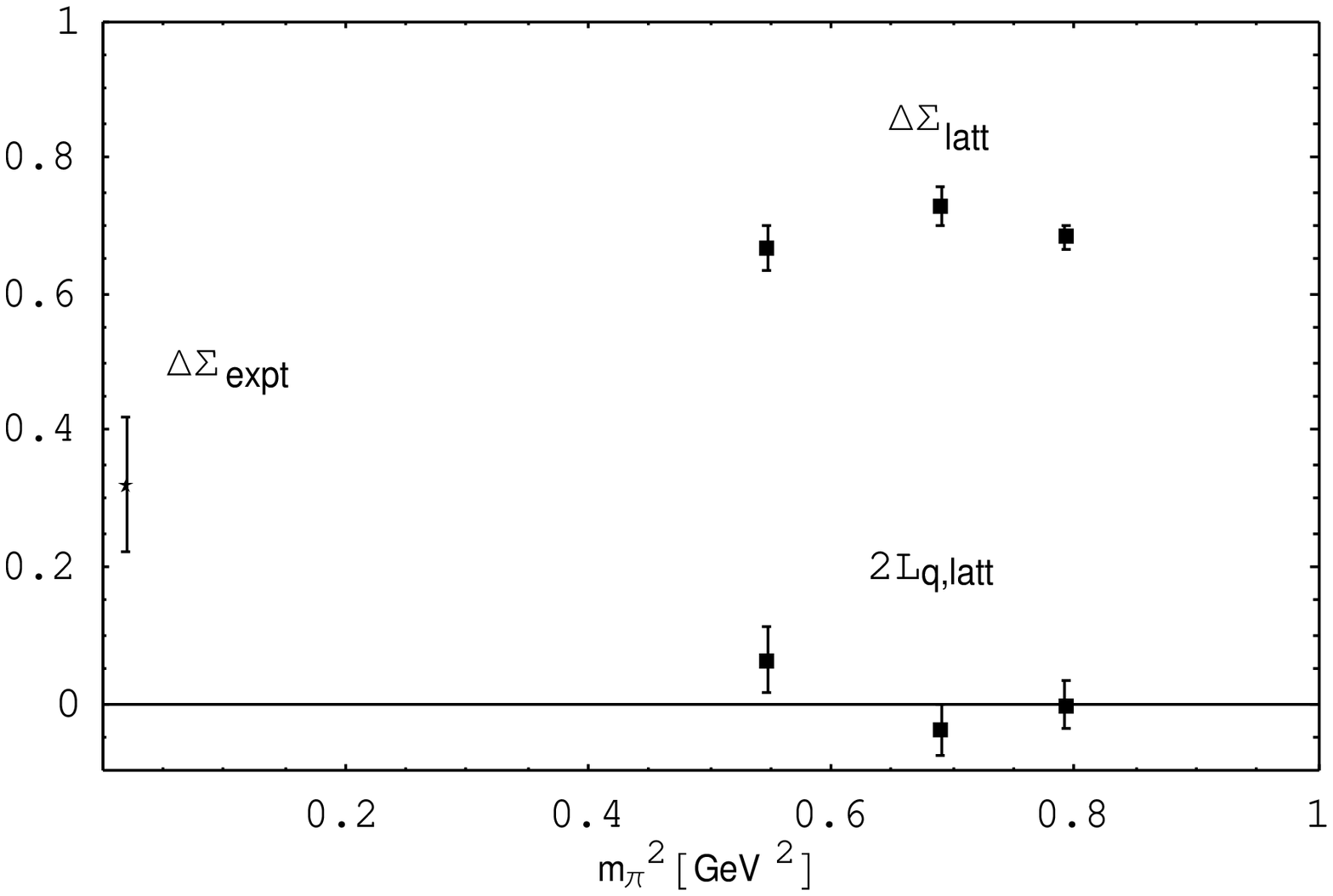}} }
  \vspace*{-0.8cm}
 \caption{Generalized form factors $A_{20} , B_{20}$ determining the total quark  contribution to the nucleon spin (left panel) and  the fraction of the nucleon spin arising from quark spins, $\Delta \Sigma$, and quark orbital angular momentum, $2 L_q$, (right panel).  }
 \label{fig:spin}
 \end{center}
   \vspace*{-.3cm}
\end{figure}

The total quark contribution to the nucleon spin\cite{Ji:1996ek} is given by the extrapolation to $t=0$ of  $A^{\mbox{\tiny u+d}}_{20}(t)$ and $B^{\mbox{\tiny  u+d}}_{20}(t)$ shown in Figure~\ref{fig:spin}.  Since  $A^{\mbox{\tiny u+d}}_{20}(t)$ is calculated directly at   $t=0$  and  $B^{\mbox{\tiny  u+d}}_{20}(t)$ is well fit by a constant that is measured to be nearly zero with small errors\cite{Hagler:2003jd}, the connected contribution to the angular momentum is measured to within a few percent. 
Combined with the calculation\cite{Dolgov:2002zm} of $\Sigma$, we obtain the connected diagram contributions to the decomposition of nucleon spin shown in the right hand portion of  Fig.~\ref{fig:spin}. Similar results have been obtained in Refs.  \raisebox{-.12cm}{\cite{Gockeler:2003jf,Mathur:1999uf}}.   To the extent that the disconnected diagrams, which have not yet been calculated, do not change the qualitative behavior, we conclude that of the order of 70\% of the spin of the nucleon arises from the quark spin and a negligible fraction arises from the quark orbital angular momentum in a heavy pion world.  With new hybrid calculations analogous to those shown for $g_a$, it will be interesting to  observe the quark spin contribution decrease to  the experimental value $\sim$ 30\%.

\smallskip
\noindent {\bf Transverse structure}\quad
Whereas structure functions measured in deep inelastic scattering only tell us about the distribution of quarks as a function of the longitudinal momentum fraction, $q(x)$, generalized parton distributions explore the quark distribution in three dimensions, $q(x, r_{\perp})$, as a function of the longitudinal momentum fraction $x$ and the transverse spatial coordinate $\vec r_\perp$. Indeed, Burkardt has shown\cite{Burkardt:2000za}  that the quantity corresponding to the mass in a conventional form factor approaches infinity for the transverse form factor in the infinite momentum frame, yielding the familiar non-relativistic relation that
the generalized parton
distribution $H(x, 0, t)$ is the transverse Fourier transform of the quark distribution, \\
$
H(x, 0, -\Delta_\perp^2) = \int d^2 r_\perp q(x, r_{\perp})\,
e^{  i \vec r_\perp \cdot  \vec \Delta_\perp}
$.
Hence, the generalized form factor, which can be calculated on the lattice, measures moments of $q(x, r_{\perp})$,
$
A_{n,0}( -\Delta_\perp^2)  
  = \int d^2 r_\perp \int dx\, x^{n-1}  q(x, r_{\perp})\,e^{
 i \vec r_\perp \cdot  \vec \Delta_\perp} .
$

\begin{figure}[t]
\begin{center}
 \hspace*{-.6cm} \hbox{ \includegraphics[scale=0.60,clip=true,angle=0]{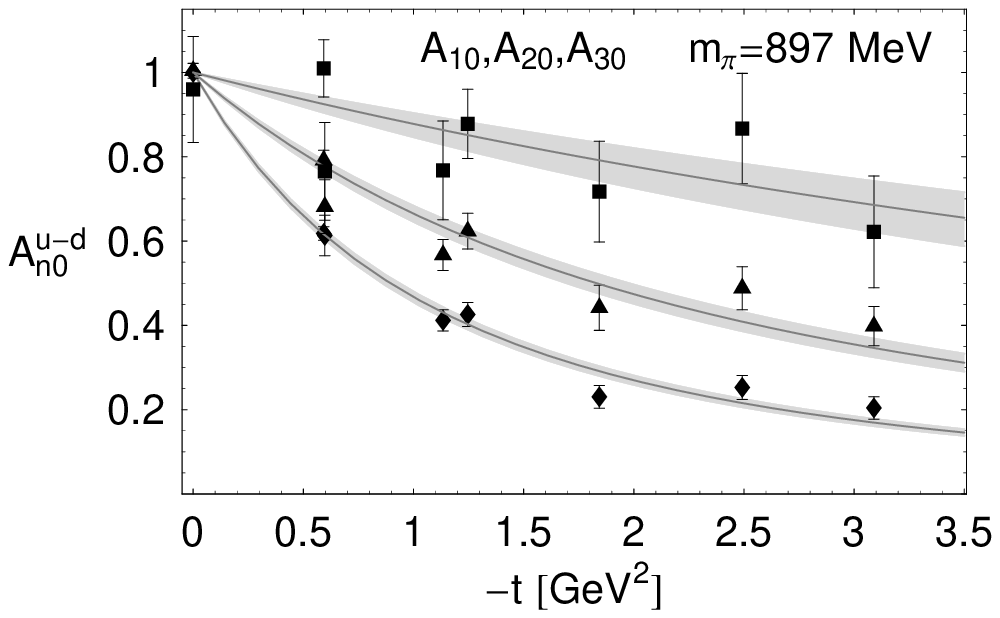} \hspace*{-.2cm}\raisebox{-.22cm}{
\includegraphics[scale=0.23,clip=true,angle=0]{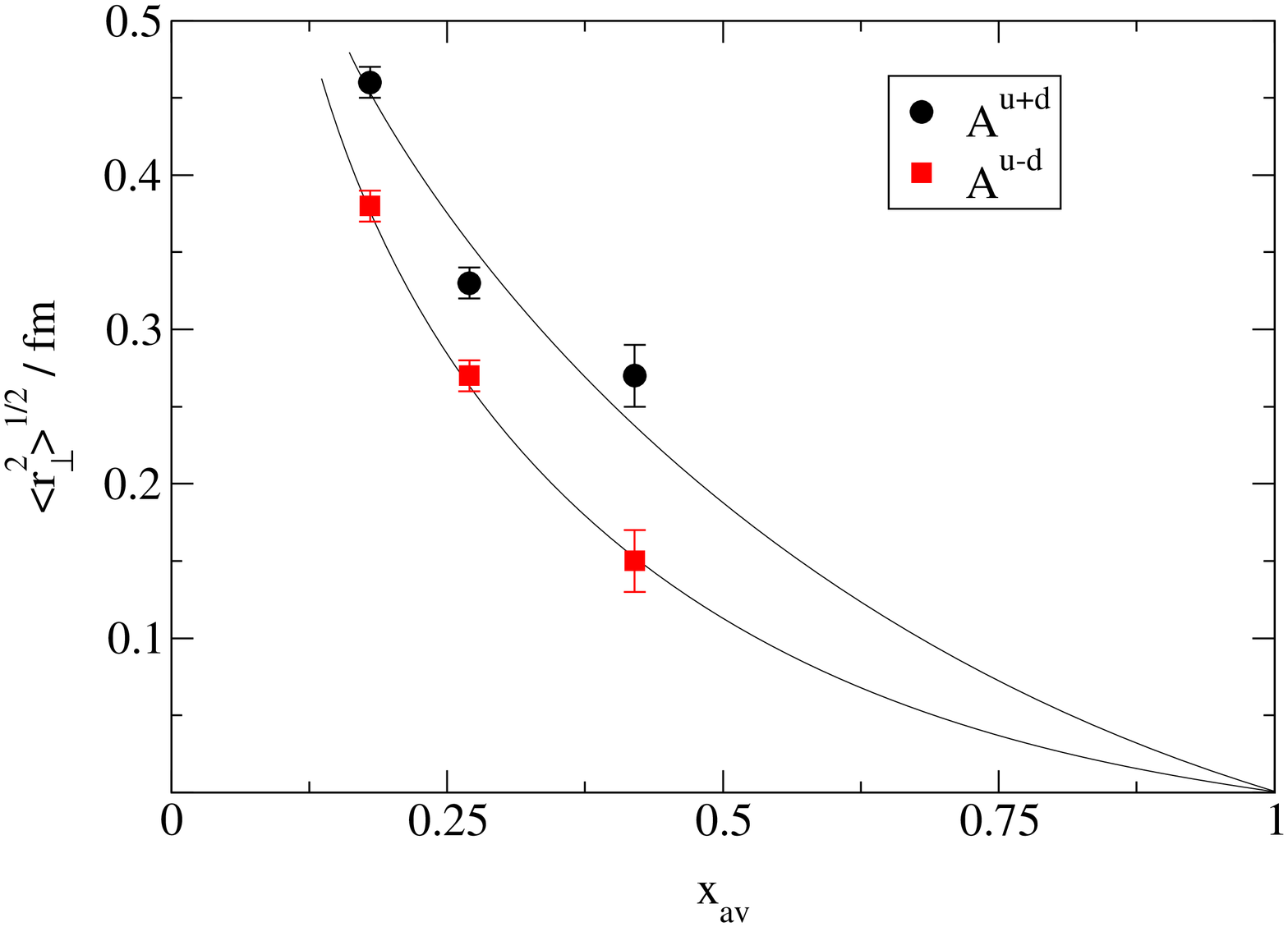}}}
\vspace*{-0.1cm}
\caption{The left panel shows normalized generalized form factors 
	 $A^{u-d}_{n,0}(t)$ 
	 for n=1 (diamonds), n=2 (triangles), and n=3 (squares). The right panel shows the transverse rms radius of the proton light cone wave function as a function of the average quark momentum fraction, $x_{av}$, for each measured moment.}
 \label{fig:trans}
 \end{center}
   \vspace*{-.5cm}
\end{figure}

Lattice results for generalized form factors in the heavy pion world have been discussed\cite{Negele:2004iu,Hagler:2003jd,:2003is,Gockeler:2004vx} in some detail, but the key features are shown in Fig.~\ref{fig:trans}. Physically, as $x \to 1$, the struck quark carries all the momentum, the spectator partons contribute negligibly, the transverse distribution approaches a delta-function $\delta(r_\perp)$, and the slope of the form factor therefore approaches zero. As $x$ decreases, successively more spectator partons are relevant, the transverse sizes increases and the slope of the form factor correspondingly increases. Hence, we expect that as $n$ increases, the moment $x^{n-1}$ increasingly weights large $x$ thereby decreasing the slope.  The left panel of Fig.~\ref{fig:trans} shows that this change in slope is quite dramatic, with the third moment being far flatter than the first moment.  

It is useful to use the slope of the form factors at $t$ = 0
to determine the transverse rms radius for each moment
$
\langle r_\perp^2 \rangle^{(n)} = {\frac{\int d^2r_\perp r^2_\perp
    \int dx\, x^{n-1} q(x,r_\perp)}{\int d^2r_\perp  \int dx\, x^{n-1}
    q(x,r_\perp)} }$,
and to plot the rms radius as a function of the average value of $x$  for that moment, as shown in the right panel of Fig.~\ref{fig:trans}. The $x$ dependence of this figure is quite striking, with the nonsinglet transverse size dropping 62\% as the mean value of $x$ increases from 0.2 to 0.4, and shows the ability of these lattice calculations to reveal significant transverse structure in light-cone wave functions.

\smallskip
\noindent {\bf Nucleon-Delta Transition}

\begin{figure}[t]
\begin{center}
\hspace*{-.4 cm} \hbox{  \includegraphics[scale=0.30,clip=true,angle=0]{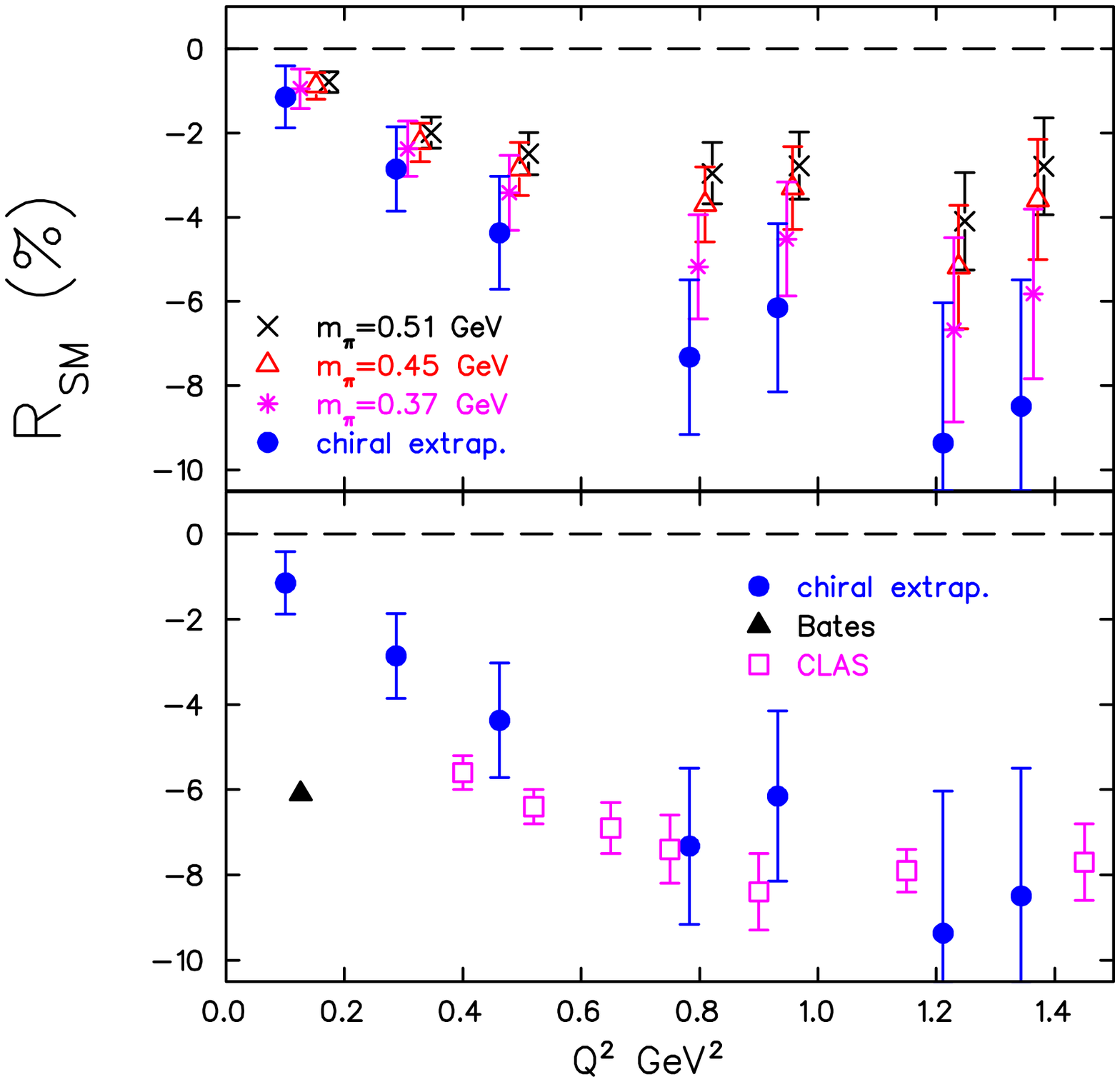}
  \hspace*{.4 cm}  \includegraphics[scale=0.30,clip=true,angle=0]{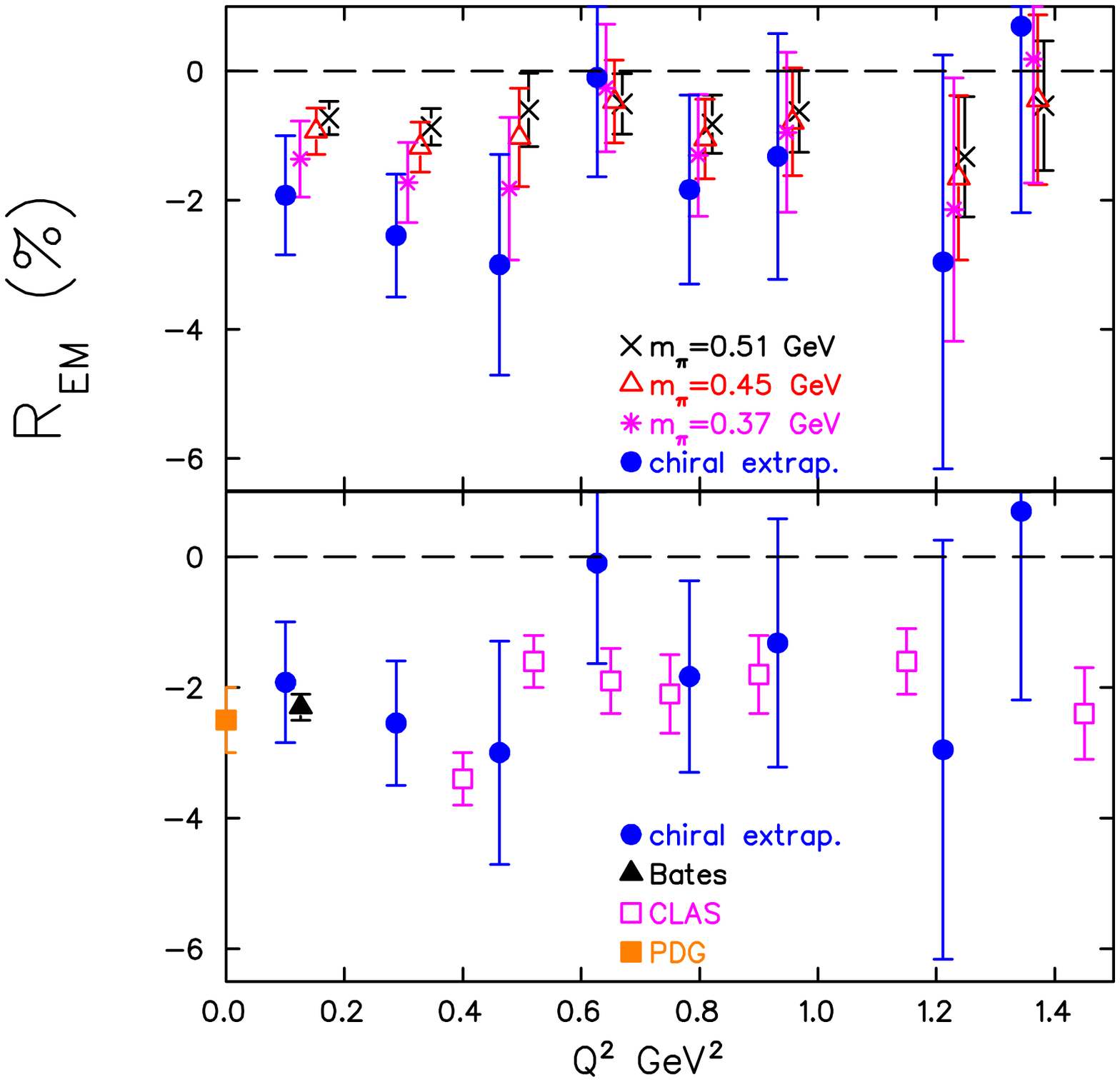}}
 \caption{Ratio of Coulomb to magnetic form factors, $R_{SM}$ (left panel) and of electric to magnetic form factors, $R_{EM}$. The upper curves show quenched calculations at several pion masses and their chiral extrapolation, and the lower curves compare the extrapolated result with experimental data.}
 \label{fig:nDelta}
 \end{center}
  \vspace*{-.3cm}
\end{figure}

Since deformed nuclei play such a prominent role in nuclear physics, it is interesting to explore whether deformation plays a comparable role in hadron structure.  The experimental method of choice to reveal the presence of deformation in the low-lying baryons is measuring the N - $\Delta$ transition amplitude, where the dominant transition is magnetic dipole (M1)  and non-vanishing electric quadrupole (E2) and Coulomb quadrupole (C2) amplitudes are a signature of deformation in the nucleon,  Delta, or both.  It is convenient to measure the ratio of the electric to magnetic form factors, $R_{EM} = -{\cal G}_{E2}(q^2) / {\cal G}_{M1}(q^2)$ and of the Coulomb to magnetic form factors, $R_{SM} = - |\vec q|{\cal G}_{C2}(q^2) / 2 m_{\Delta} {\cal G}_{M1}(q^2)$.
Figure~\ref{fig:nDelta} shows the results of a new lattice method that for the first time has the precision to measure non-vanishing $R_{EM}$ and $R_{SM}$ ratios\cite{Alexandrou:2004xn,Alexandrou:2003ea}. Extrapolation of these quenched results in the heavy pion regime to the chiral limit yields results qualitatively similar to experiment, raising the expectation that calculations presently under way in the chiral regime will provide quantitative agreement with experiment. With this absolute calibration from experiment, comparable lattice calculations of correlation functions in the $\Delta$ can then be used to obtain insight into the magnitude and origin of its deformation. 

\section{New Era of Lattice QCD}

  From these brief examples, I hope it is clear to experimentalists that lattice QCD is finally becoming a quantitative tool that can be expected to agree with experiment and to complement it where experiments are impractical. When the credibility of {\it ab initio} calculations has been well established by agreement with experiment, I hope it is also clear to theorists that lattice calculations will also enable a host of illuminating calculations of the internal structure of hadrons.  A case in point is  the exploration of diquark components in hadrons and calculation of  pentaquarks,  to which I could not do justice in this short summary.
  
 \smallskip
\noindent {\bf Acknowledgements}  The work described in this talk was a collaborative effort arising from the work of many authors cited in the references. I particularly wish to acknowledge the major contributions of Constantia Alexandrou, Philipp H\"agler,  Kostas Orginos, Dru Renner, Wolfram Schroers, and Antonios Tsapalis to the work I discussed.
We are also  indebted to members of the MILC  and SESAM collaborations for the dynamical quark configurations which made our full QCD calculations possible.

\bibliography{GDH_References}

\end{document}